\documentclass[journal,letterpaper]{IEEEtran}

\usepackage{amsmath,amsfonts}
\usepackage{algorithm,algorithmic}
\usepackage{array}
\usepackage[colorlinks,linkcolor=blue,anchorcolor=blue,citecolor=blue]{hyperref}
\usepackage[caption=false,font=normalsize,labelfont=sf,textfont=sf]{subfig}
\usepackage{textcomp}
\usepackage{stfloats}
\usepackage{url}
\usepackage{verbatim}
\usepackage{graphicx}
\usepackage{cite}
\usepackage{booktabs}
\usepackage{balance}
\usepackage{multirow}
\usepackage[normalem]{ulem}
\usepackage{xcolor}
\usepackage{booktabs}
\usepackage{multirow}
\usepackage{arydshln}

\useunder{\uline}{\ul}{}
\begin{document}
\title{A H.265/HEVC Video Steganalysis Algorithm Based on CU Block Structure Gradients and IPM Mapping}
\author{Xiang Zhang, Haiyang Xia, Fan Wang, Ziwen He, Wenbin Huang, Fei Peng, Zhangjie Fu*

\thanks{This work was supported in part by the National Natural Science Foundation of China under Grant 62372128, 62401270, 62502215, U22B2062, 62172232; China Postdoctoral Science Foundation under Grant 2023M741778. (\textit{Corresponding author: Zhangjie Fu})

Xiang Zhang, Haiyang Xia, Ziwen He, Wenbin Huang, and Zhangjie Fu are with the Engineering Research Center of Digital Forensics, Ministry of Education, Nanjing University of Information Science and Technology, Nanjing, Jiangsu 210044, China (e-mail: zhangxiang@nuist.edu.cn; 202412490784@nuist.edu.cn; ziwen.he@nuist.edu.cn; wenbinhuang@nuist.edu.cn; fzj@nuist.edu.cn).

Fan Wang is with the Faculty of Science and Technology, University of Macau, Macau, 999078, China (e-mail: wangfan@um.edu.mo).

Fei Peng is with the School of Artificial Intelligence, Guangzhou University, Guangzhou, Guangdong 510006, China (e-mail: eepengf@gmail.com).
}}
\markboth{Journal of \LaTeX\ Class Files,~Vol.~14, No.~8, August~2021}%
{Shell \MakeLowercase{\textit{et al.}}: A Sample Article Using IEEEtran.cls for IEEE Journals}


\maketitle

\begin{abstract}
Existing H.265/HEVC video steganalysis research mainly focuses on detecting the steganography based on motion vectors, intra prediction modes, and transform coefficients. However, there is still a lack of steganalysis methods that effectively detect steganography based on Coding Unit (CU) block structure of I frames. This makes such steganographic methods highly evasive to existing detection systems, thereby potentially posing significant risks to information security and the regulation of covert communications. To address this issue, we propose, for the first time, a H.265/HEVC video steganalysis algorithm based on CU block structure gradients and Intra Prediction Mode (IPM) mapping. Firstly, we construct a gradient map to describe the changes in the CU block structure due to steganography, and combines it with a pixel-level mapping representation of IPM. It can jointly model the multiple structural perturbations introduced by steganography based on CU block structure. Then, we design a novel steganalysis network, namely GradIPMFormer to effectively improve the perception ability of structural perturbations of CU blocks. In GradIPMFormer, convolutional local embedding is integrated with token modeling based on Transformer to separately capture the intra-block local boundary perturbations and long-range inter-block structural dependencies. Experimental results show that under different quantization parameters and resolution settings, the proposed method consistently achieves superior detection performance across multiple steganography methods based on CU block structure. This study provides a new H.265/HEVC steganalysis paradigm for detecting CU block structure-based steganography methods, thus having significant research value for promoting the detection of covert communication.
\end{abstract}


\section{Introduction}\label{intro}
\IEEEPARstart{W}{ith} the rapid development of multimedia technologies, massive amounts of information are transmitted over networks. Video steganography is a technique that enables covert transmission by hiding secret information in videos, and it is widely used in fields such as national defense and the military~\cite{tew2013overview,xie2026large}. However, the abuse of this technology by criminals can pose serious threats to public security. Therefore, video steganalysis has emerged, which can effectively detect whether videos transmitted through a channel contain secret information, thus preventing hidden malicious information from being transmitted. In general, video steganalysis needs to be integrated with video decoding technologies. As one of the mainstream encoding and decoding standards, H.265/HEVC offers significant advantages over the previous H.264/AVC standard, including higher compression efficiency, support for higher resolutions, and better network adaptability~\cite{sullivan2012overview,ohm2012comparison}. Therefore, H.265/HEVC-based video steganalysis has become a major research focus. Existing H.265/HEVC-based video steganalysis methods can be categorized according to the type of carrier being analyzed: inter-frame information-based steganalysis, transform residual coefficient-based steganalysis, and intra prediction mode-based steganalysis~\cite{tan2016dense}.

Inter-frame information-based steganalysis aims to expose stego videos by characterizing irregularities in temporal syntax elements. Early works often exploited partition-level cues. For example, Li et al.~\cite{li2019hevc} summarized PU partition-mode statistics to construct discriminative feature vectors for classification, while Dai et al.~\cite{dai2023hevc} transformed each frame into a PU partition map and employed a multi-scale residual network to learn cross-scale evidence from such structure-aware representations. Beyond partition modes, Liu et al.~\cite{liu2021hevc} investigated the behavior of motion vectors (MVs) by measuring the rate–distortion discrepancies between neighboring and candidate MVs, thereby forming an explicit feature set for detection. Zhai et al.~\cite{zhai2019universal} proposed a universal video steganalysis method based on motion vector consistency. It designed a 12-dimensional feature set for multi-domain steganalysis and achieved strong detection performance in both matching and non-matching domain scenarios. These approaches indicate that temporal syntax elements (e.g., partition modes and MVs) can provide useful forensic cues, although their effectiveness may depend on the reliability of temporal modeling and the subtlety of the embedding strategy.

In transform residual coefficient-based steganalysis, Wang et al.~\cite{wang2017steganalytic} combined intra-frame residual statistics with inter-frame temporal descriptors to enhance discriminability under transform-coefficient perturbations. Zhang et al.~\cite{zhang2020video} further demonstrated that coefficient embedding can influence the deblocking process and designed a high-dimensional feature set based on luminance modification patterns. More recently, Dai et al.~\cite{dai2025hevc} revealed that distortion compensation embeddings lead to concentrated error patterns in coefficients, and introduced a prediction-error map together with attention-enhanced deep modeling to improve sensitivity at low payloads. Overall, transform residual coefficient-based steganalysis can be effective to some extent. However, its performance is easily affected by the high dependence of residual statistics on coding parameters and scene content. 

Intra prediction mode (IPM)-based steganalysis exploits the fact that embedding constraints alter intra prediction decisions, thereby constructing detectable statistical artifacts in the distribution of intra prediction modes. Zhao et al.~\cite{zhao2015video} proposed IPM calibration features by measuring IPM shift tendencies and SATD-related variations. Subsequently, Sheng et al.~\cite{sheng2017prediction} proposed a steganalysis method based on the structure of intra prediction units. Specifically, a six-dimensional feature unit is constructed by calculating the differences in the number and proportion of prediction units before and after re-compression, thereby detecting the structural disruptions caused by steganography. Following this idea, Liu et al.~\cite{liu2020steganalysis} developed an end-to-end network based on high-pass filtering and residual learning to amplifying the steganograohic traces in  reconstructed frames. Although IPM-based steganalysis can directly reflect disturbances in the prediction process, they still struggle to capture structure-coupled changes when steganographic embedding mainly manipulates CU partition structures rather than directly modifying individual IPM values. 

In summary, existing steganalysis has been developed based on the syntax elements mentioned above. However, steganalysis targeting CU block structure-based steganography remains relatively unexplored. CU block structure-based steganography focuses on modifying the CU block structure of I-frames to embed information. Such modifications inevitably undermine the original optimality of CU partition \cite{tew2014information,dong2022adaptive,yang2024quad,wang2024adaptive}. The destruction mainly reflects three aspects, including the change of block structure, the reorganization of hierarchical relationship and the discontinuity of structure. However, existing studies typically evaluate steganalysis resistance using detectors that are not specifically designed for CU block structure-based steganography. For example, Yang et al.~\cite{yang2024quad} employed two intra prediction mode-based steganalysis methods \cite{zhao2015video,sheng2017prediction} and two inter-frame information-based steganalysis methods \cite{zhai2019universal,dai2023hevc} to assess the security of their algorithm. Such steganalysis is insufficient to effectively measure the true security of these algorithms, which has become an important factor limiting the development of CU block structure-based steganography. Therefore, there is an urgent need for a dedicated and effective steganalysis method to detect CU block structure-based steganography. However, how to precisely characterize steganographic behaviors from the perspective of the coding structure, and build steganalysis model that is sensitive to structural perturbations while maintaining strong generalization ability, is a challenging issue.


To address the above issue, this paper proposes a video steganalysis algorithm based on CU block structure gradients and IPM mapping. 
Unlike previous steganalysis that statistically analyze the structural variations of that syntax element before and after embedding, we observe that the structural differences introduced by steganographic embedding based on the CU block structure are often subtle.
As a result, 
such statistical analysis methods not only suffer significant performance degradation but  even  become ineffective when detecting steganographic schemes with low embedding rates and high
imperceptibility.
\begin{figure}[htbp]
    \centering
    \includegraphics[width=1\linewidth]{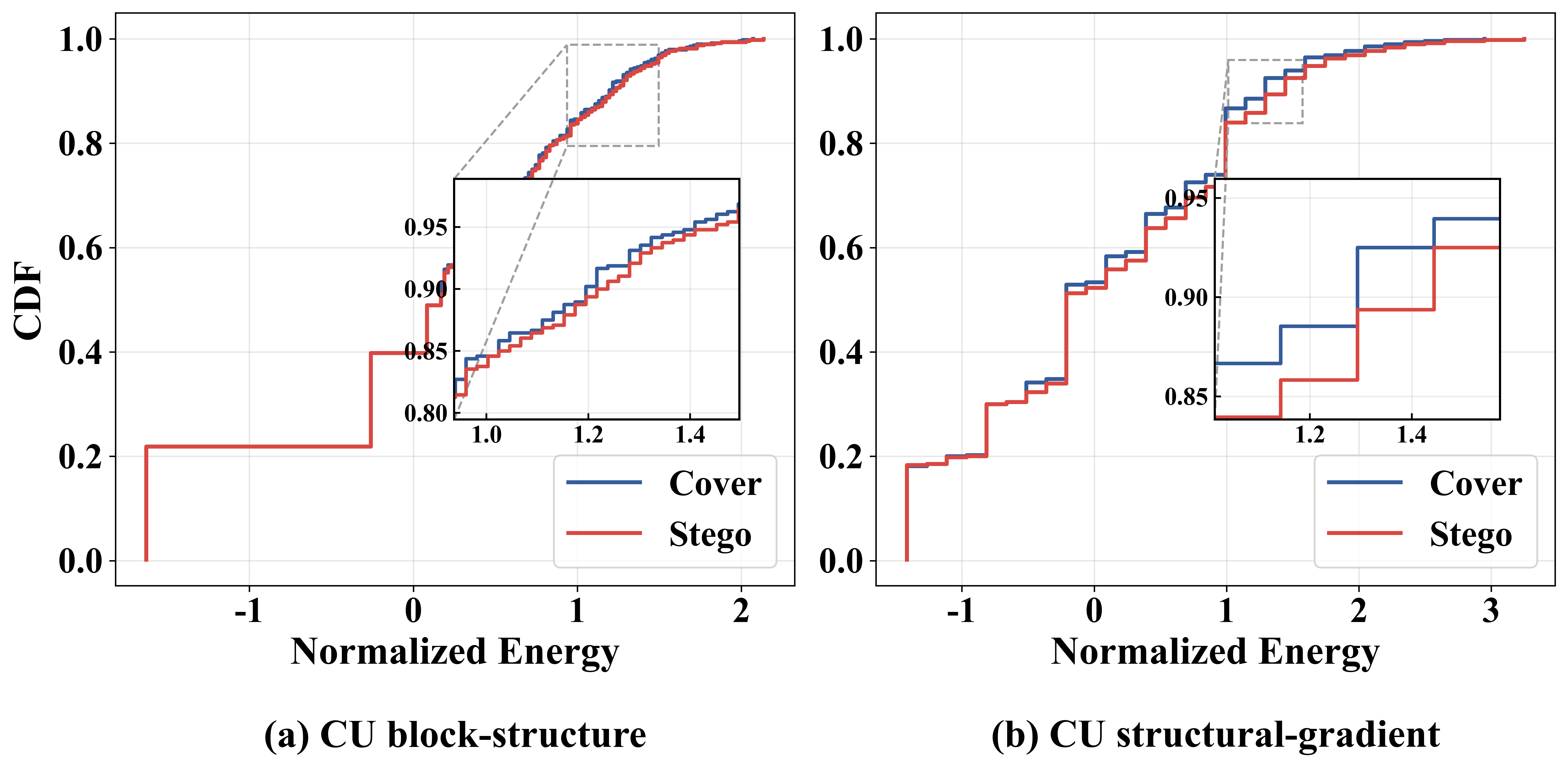}
    \caption{Comparison of CDFs for CU block-structure map and CU structural-gradient map}
    \label{CDF}
\end{figure}

To demonstrate this, we generate the stego video of ``Aspen'' sequence using the method proposed by Dong et al.~\cite{dong2022adaptive}. We then choose frame \#1 of the cover and stego video, partition the CU block-structure map into non-overlapping $16\times16$ patches. Afterwards, we compute the energy of each patch as $e=\frac{1}{|B|}\sum_{(i,j)\in B}|x_{ij}|$, where $B$ denotes a local $16\times16$ patch of the map, $|B|$ is the number of pixels within the patch, and $x_{ij}$ represents the pixel value at spatial position $(i,j)$ within $B$. The obtained energies are further $z$-score normalized with respect to the cover statistics, and we plot the empirical CDF, which is shown in Fig.~\ref{CDF}(a). It can be observed that the cumulative distributions of CU block structures in the cover and stego videos largely overlap. This indicates that relying solely on CU block structure statistics is insufficient to effectively distinguish between cover and stego videos. In contrast, we observe that \textit{when \textbf{gradient modeling} is further incorporated into the CU block structure, the cumulative distribution changes significantly.} As illustrated in Fig.~\ref{CDF}(b), we compute the cumulative distributions of the CU structural-gradient maps using the same procedure. compared with the cover video, the stego video exhibits a more pronounced shift in the overall gradient distribution. This phenomenon provides key inspiration for the design of our proposed method.

\begin{figure}[htbp]
    \centering
    \includegraphics[width=0.8\linewidth]{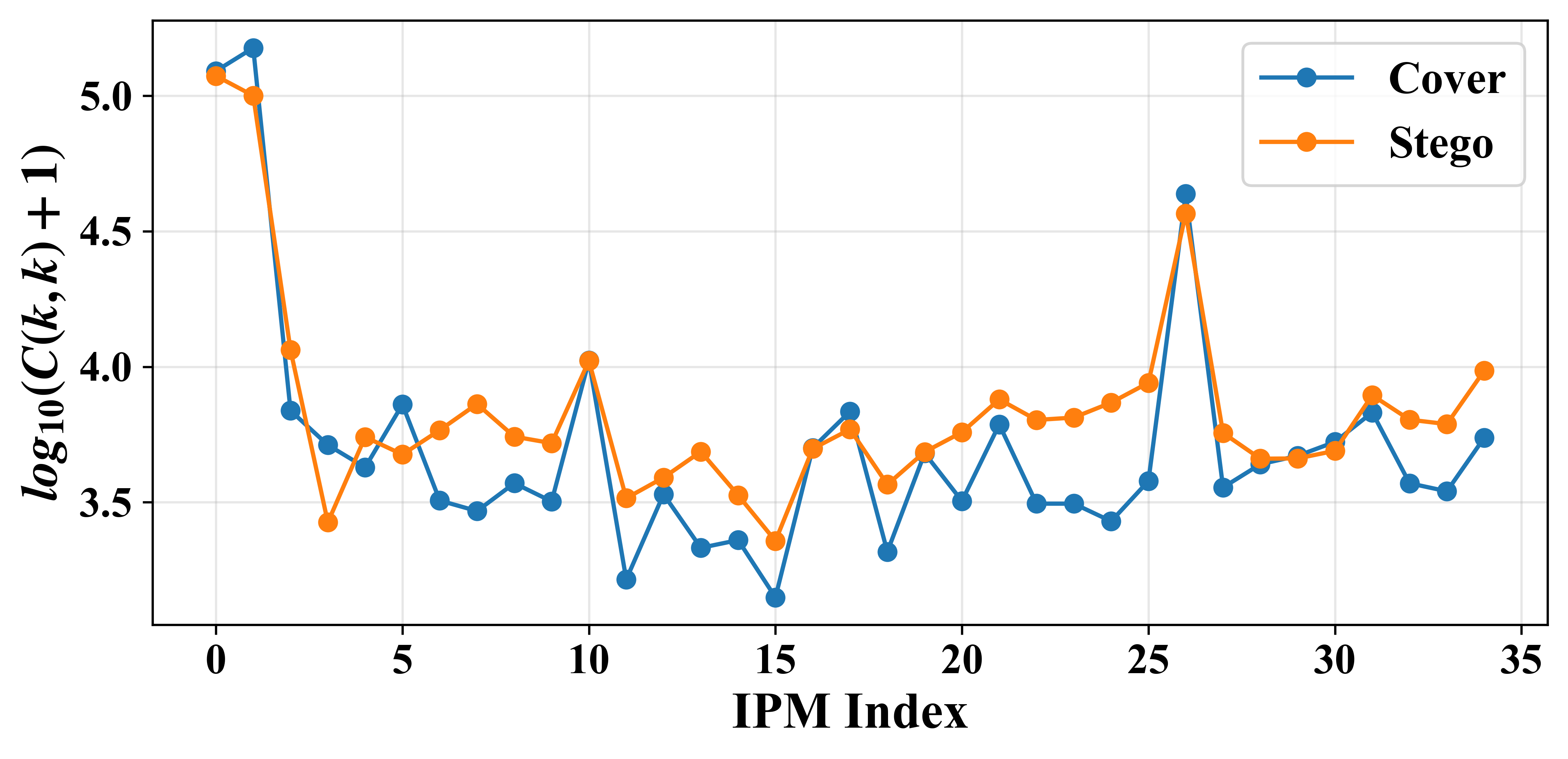}
    \caption{Differences in IPM co-occurrence matrix distributions before and after steganography}
    \label{IPM Change}
\end{figure}

Moreover, through an in-depth investigation of the relationship between CU block structures and intra prediction modes, we further observe an important phenomenon: \textit{when the CU block structure is modified, the optimal intra prediction mode (IPM) of the corresponding block also changes accordingly. We refer to this intriguing phenomenon as \textbf{IPM drift phenomenon.}} To illustrate this phenomenon, we apply the steganography method proposed by Dong et al.~\cite{dong2022adaptive} to ``Aspen'' sequence, and compute the IPM co-occurrence matrices for both the cover and stego videos. Fig.~\ref{IPM Change} shows the diagonal statistics of the IPM co-occurrence matrix for cover and stego videos, plotted as log form. 
From the figure, it can be observed that the consistent gap between the curves before and after the steganographic embedding verify that the embedding operation has disrupted the local consistency of IPMs. This provides intuitive and powerful evidence for the IPM drift phenomenon. 
From the coding mechanism perspective, CU partition decisions and IPM
decisions are not independent. Once the CU partition structure is modified,
the spatial support of the prediction block, the available neighboring
reference samples, and the residual distribution used in SATD/RDO
evaluation may also change. Consequently, the ranking of candidate intra
prediction modes can be affected, leading to structure-induced IPM drift.
This mechanism explains why CU block structure steganography may leave
not only partition-level traces but also prediction-mode-level traces,
thereby motivating the joint modeling of CU structure gradients and IPM
mapping in our framework.

Based on the above two observations, we construct the complete steganalysis framework proposed in this paper. Specifically, we first generate a CU block structure gradient map to capture the spatial partition boundaries and hierarchical variations of coding units. Meanwhile, we introduce a pixel-level mapping representation of IPM to model their distribution characteristics under CU structural constraints. By aligning and fusing the structural gradient information with the IPM mapping representation, the proposed method is able to more comprehensively characterize the abnormal patterns introduced by CU block-based steganographic embedding. Moreover, we develop a dedicated network for CU block structure steganalysis, termed GradIPMFormer. The network is built upon a Transformer architecture and performs global modeling of CU structural gradients and IPM mappings through sequence representation and self-attention mechanisms, thereby enhancing its capability to perceive subtle structural perturbations. In summary, the main contributions of this paper are as follows:

\begin{itemize}
\item \textbf{A novel video steganalysis specifically designed to detect CU block structure-based steganography is explored.} Unlike existing steganalysis methods that mainly focus on motion vectors, transform residual coefficients, or intra prediction modes, this work is the first to take a CU block structure perspective and, systematically investigates how CU block structure-based steganography affects CU structures and IPM decisions, providing a new research paradigm for H.265/HEVC video steganalysis.

\item \textbf{A steganalysis feature representation based on block structure gradients and IPM mapping is constructed.} We propose a novel block structure gradient mapping and comprehensively analayze the influence of CU block structure on the spatial distribution of IPM  
before and after steganographicembedding, 
thereby overcoming the limitations of relying on a single syntax element.

\item \textbf{A new network for CU block structure steganalysis, termed GradIPMFormer, is proposed.} To the best of our knowledge, this is the first work to introduce Transformer architecture into video steganalysis. By jointly modeling the spatial dependencies of CU block structure variations and their syntax correlations, 
the proposed GradIPMFormer combines convolutional feature extraction with the self-attention mechanism to more effectively capture the subtle perturbations introduced by the CU block structure-based steganography.

\item \textbf{Extensive experiments are conducted to validate the effectiveness of the proposed framework.} 
Experimental results across different quantization parameters, resolutions, network architectures, and multiple CU block structure-based steganography algorithms show that  our approach consistently achieves superior detection performance, demonstrating its feasibility and effectiveness.
\end{itemize}

The remainder of this paper is organized as follows: Section \ref{sec2} reviews preliminaries. Section \ref{sec3} describes the proposed steganalysis model. Section \ref{sec4} discusses the experimental results and analysis. Finally, Section \ref{sec5} is the conclusion.

\section{Preliminaries}\label{sec2}
\subsection{Principles of Coding Unit in H.265/HEVC}\label{II-A}

In H.265/HEVC intra coding, each frame is first partitioned into Coding Tree Units (CTUs), typically with a size of $64\times64$. Within each CTU, a quadtree-based recursive partitioning structure is adopted to generate Coding Units (CUs) at different depth levels. Let the quadtree depth index be denoted as $Dh$, where $Dh=0$ corresponds to the CTU level.
\begin{figure}[t]
    \centering
    \includegraphics[width=1\linewidth]{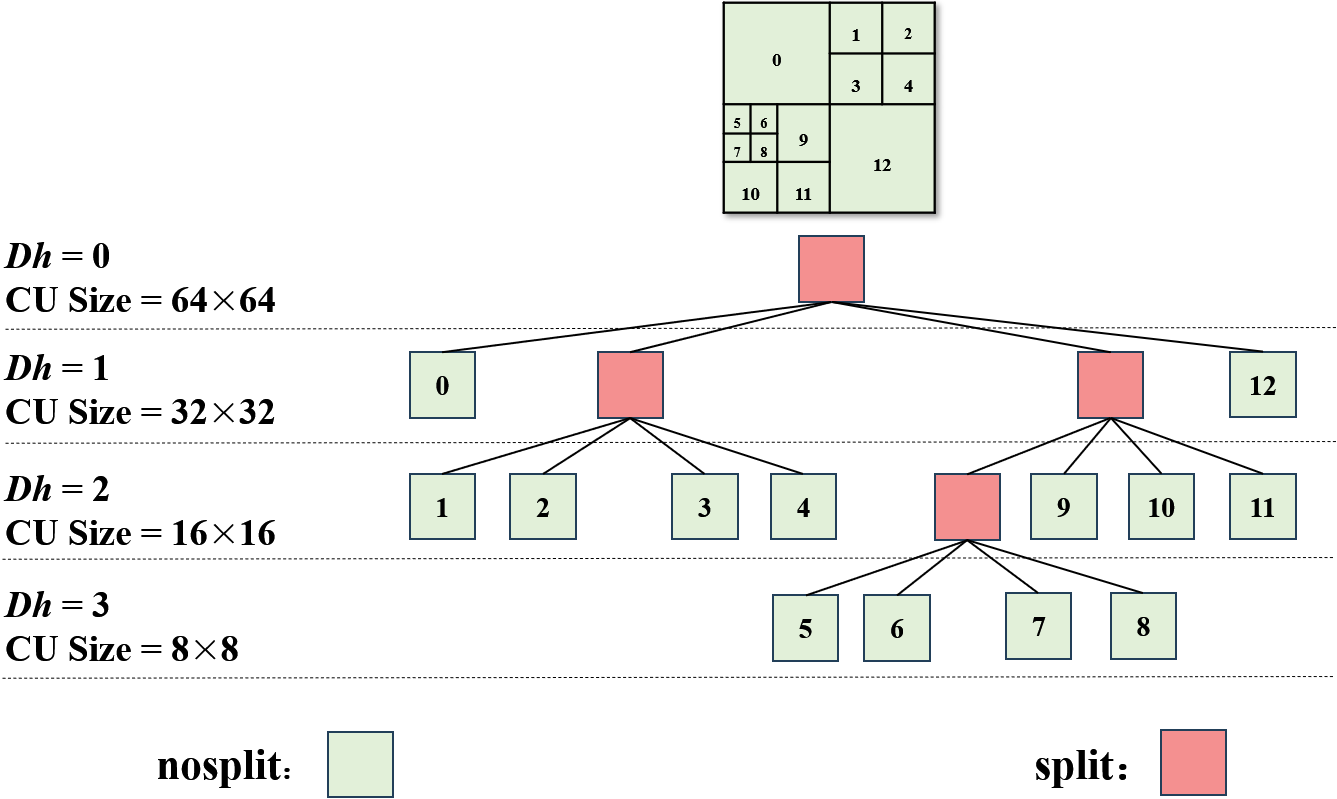}
    \caption{CU quadtree partition structure in H.265/HEVC. }
    \label{fig:CU_quadtree}
\end{figure}
As illustrated in Fig.~\ref{fig:CU_quadtree}, the quadtree partition proceeds in a top-down manner, while the RDO decision is finalized in a bottom-up fashion. At each depth $Dh$, a CU of size $2^{-Dh}\!\times$CTU may either terminate the partitioning process or be further split into four equally sized sub-CUs at depth $Dh+1$, until a predefined minimum CU size is reached. The CU partition decision is governed by the rate--distortion optimization (RDO) mechanism. For a CU at depth $Dh$, two competing coding configurations are evaluated:

\textbf{1) Non-split configuration.}  
When the CU is not split, intra prediction is performed within this block. Multiple candidate intra prediction modes are first evaluated using a fast cost metric, and a reduced candidate set is selected for full rate--distortion optimization. For each candidate mode $m$, the RD cost is computed as:
\begin{equation}
J(m) = D(m) + \lambda R(m),
\label{eq:RD_mode}
\end{equation}
where $D(m)$ denotes the distortion between the original block and the reconstructed block under mode $m$, $R(m)$ denotes the total number of bits required to encode the intra prediction mode $m$ and associated syntax elements, and $\lambda$ is the Lagrange multiplier determined by the quantization parameter (QP). The optimal intra mode $m^{*}$ is selected by minimizing $J(m)$, and non-split RD cost of the CU is:
\begin{equation}
J_{\mathrm{NS}} = J(m^{*}) + \lambda R_{\mathrm{splitflag}=0},
\label{eq:J_NS_detailed}
\end{equation}
where $R_{\mathrm{splitflag}=0}$ denotes the signal cost of no further partitioned of the CU.

\textbf{2) Split configuration.}  
If the CU is split, the encoder recursively evaluates each of the four sub-CUs at depth $Dh+1$. Let $J_k$ denote the final RD cost of the $k$-th sub-CU after its own partitioning decision is completed. The total split cost is:
\begin{equation}
J_{\mathrm{S}} = \sum_{k=1}^{4} J_k + \lambda R_{\mathrm{splitflag}=1},
\label{eq:J_S_detailed}
\end{equation}
where $R_{\mathrm{splitflag}=1}$ represents the signal cost of the current CU being further partitioned.

\textbf{3) Partition decision.}  
The final decision for the current CU is obtained as:
\begin{equation}
J^{*} = \min(J_{\mathrm{NS}}, J_{\mathrm{S}}).
\label{eq:J_compare_detailed}
\end{equation}

If $J_{\mathrm{S}} < J_{\mathrm{NS}}$, the split configuration is selected; otherwise, the CU remains unsplit.

\subsection{Principles of Intra Prediction Mode}\label{II-B}
In H.265/HEVC, intra prediction generates the predicted pixels of the current block by interpolating along a specified direction using reconstructed neighboring pixels from the top and left blocks, thereby reducing spatial redundancy. Compared with previous standards, H.265/HEVC expands the intra prediction space to a total of 35 modes, including planar (mode 0), DC (mode 1), and 33 directional modes (modes 2--34). H.265/HEVC typically adopts a two-stage fast mode decision strategy to select the optimal intra prediction mode. In the first stage, all candidate modes $m\in\{0,1,\dots,34\}$ are quickly evaluated using a SATD-based metric. Let $\mathbf{O}$ denote the original block, and $\mathbf{Y}_m$ denote the intra prediction block generated under mode $m$. The prediction residual is:
\begin{equation}
\mathbf{E}_m = \mathbf{O} - \mathbf{Y}_m.
\label{eq:residual}
\end{equation}

Then, SATD cost is computed by applying a Hadamard transform $\mathcal{H}(\cdot)$ as:
\begin{equation}
\mathrm{SATD}(m)=\sum_{i,j}\left|\mathcal{H}(\mathbf{E}_m)_{i,j}\right|.
\label{eq:satd}
\end{equation}

All the modes are ranked by $\mathrm{SATD}(m)$, and only a small set of top-ranked modes is kept as the candidate set $\mathcal{M}_{\text{cand}}$.
In the second stage, full rate--distortion optimization (RDO) is performed only on $m\in\mathcal{M}_{\text{cand}}$. Let $\mathbf{RE}_m$ denote the reconstructed block after prediction, transform, quantization, inverse transform, and reconstruction under mode $m$. The distortion term is measured by:
\begin{equation}
D_m=\|\mathbf{O}-\mathbf{RE}_m\|_2^2,
\label{eq:sse}
\end{equation}
and the Lagrangian RD cost is defined as:
\begin{equation}
J_m = D_m + \lambda R_m,
\label{eq:rdcost}
\end{equation}
where $R_m$ denotes the number of bits required to signal the mode and related syntax elements under mode $m$, and $\lambda$ is the Lagrange multiplier determined by the QP. The final optimal IPM is selected as:
\begin{equation}
m^{*}=\arg\min_{m\in\mathcal{M}_{\text{cand}}} J_m.
\label{eq:best_ipm}
\end{equation}

\subsection{Transformer Network} \label{II-c}
Transformer is a sequence modeling architecture originally proposed for machine translation, which relies entirely on self-attention mechanisms to capture global dependencies without recurrence or convolution. Compared with convolutional neural networks that primarily model local spatial correlations, Transformer is capable of establishing long-range interactions between tokens in a parallel and adaptive manner. Given an input token sequence 
\(
\mathbf{T^0} = [\mathbf{t}_1, \mathbf{t}_2, \ldots, \mathbf{t}_N] \in \mathbb{R}^{N \times d},
\)
where \(N\) denotes the number of tokens and \(d\) represents the embedding dimension, the Transformer encoder processes the sequence through stacked self-attention and feed-forward layers to obtain context-aware representations.

\subsubsection{Multi-Head Self-Attention}

The core component of Transformer is the Multi-Head Self-Attention (MHA) mechanism. 
For each token sequence \(\mathbf{T^0}\), three linear projections are first applied to obtain the query, key, and value matrices:
\begin{equation}
\mathbf{Q} = \mathbf{T^0} \mathbf{W}_q, \quad
\mathbf{K} = \mathbf{T^0} \mathbf{W}_k, \quad
\mathbf{V} = \mathbf{T^0} \mathbf{W}_v,
\end{equation}
where \(\mathbf{W}_q, \mathbf{W}_k, \mathbf{W}_v \in \mathbb{R}^{d \times d}\) are learnable projection matrices. The scaled dot-product attention is then computed as:
\begin{equation}
\text{Attention}(\mathbf{Q}, \mathbf{K}, \mathbf{V})
=
\text{Softmax}\left(
\frac{\mathbf{Q}\mathbf{K}^T}{\sqrt{d_k}}
\right)\mathbf{V},
\end{equation}
where \(d_k\) denotes the dimension of each attention head. To enhance representation capacity, multiple attention heads are employed in parallel:
\begin{equation}
\text{MHA}(\mathbf{T^0})
=
\text{Concat}(\text{head}_1, \ldots, \text{head}_h)\mathbf{W}_o,
\end{equation}
where \(h\) is the number of heads and \(\mathbf{W}_o\) is a learnable output projection matrix. This design enables the model to capture diverse relational patterns across different representations.

\subsubsection{Feed-Forward Network and Residual Structure}

Each Transformer encoder block consists of two sublayers: (i) a multi-head self-attention module and (ii) a position-wise feed-forward network (FFN). The FFN is defined as:
\begin{equation}
\text{FFN}(\mathbf{T^0})
=
\mathbf{W}_2 \sigma(\mathbf{W}_1 \mathbf{T^0}),
\end{equation}
where \(\sigma(\cdot)\) denotes a nonlinear activation function such as ReLU or GELU. To stabilize training and facilitate gradient propagation, residual connections and Layer Normalization (LN) are applied around each sublayer. 
Adopting the widely used Pre-Normalization formulation, the encoder block can be expressed as:
\begin{align}
\mathbf{T^0}' &= \mathbf{T^0} + \text{MHA}(\text{LN}(\mathbf{T^0})), \\
\mathbf{T^l} &= \mathbf{T^0}' + \text{FFN}(\text{LN}(\mathbf{T^0}')).
\end{align}

Multiple encoder blocks are stacked to construct the full Transformer encoder, enabling hierarchical feature refinement and global context modeling.

\section{The Proposed Steganalysis Model}\label{sec3}
\subsection{Overall Framework} \label{III-A}
\begin{figure*}[htbp]
    \centering
    \includegraphics[width=\linewidth]{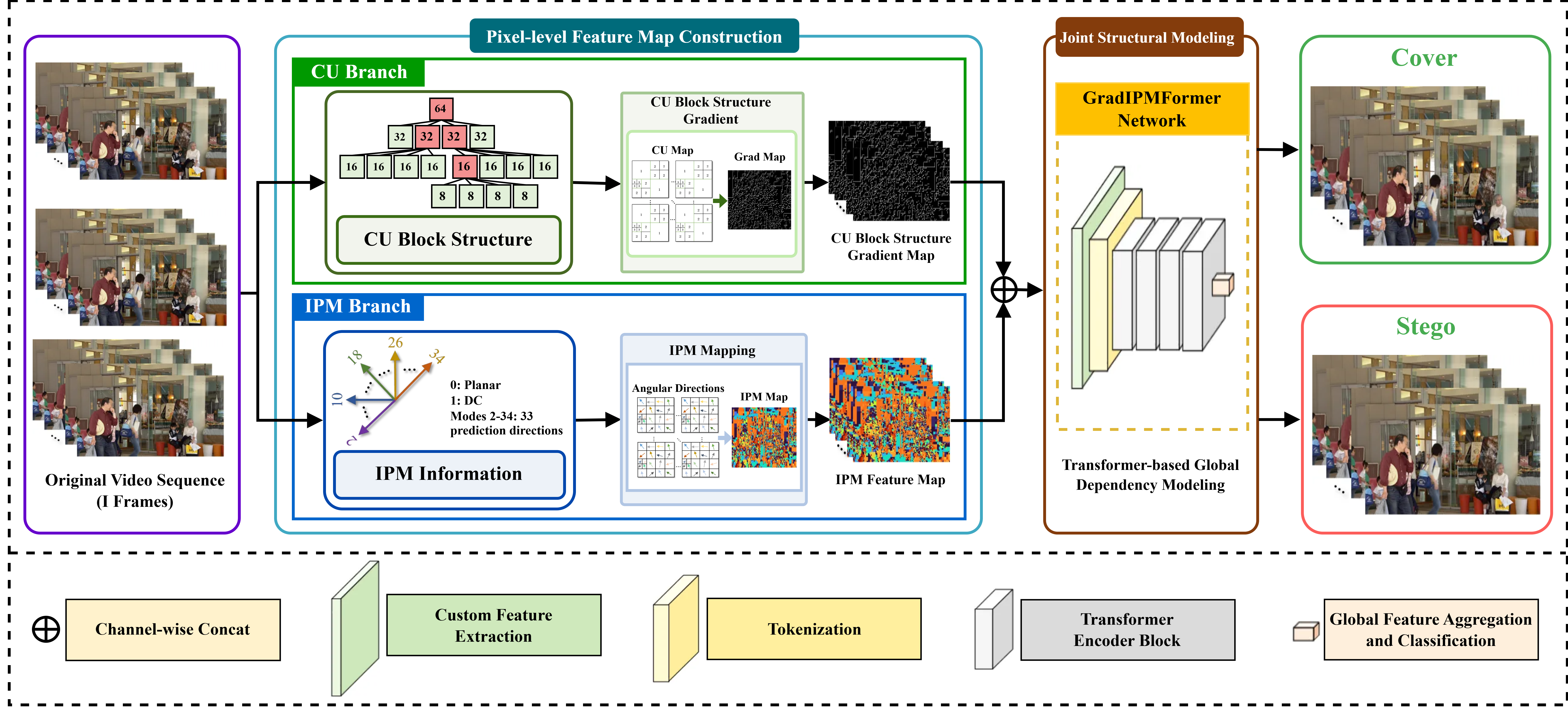}
    \caption{The proposed steganalysis framework. Given an input I frame, the CU branch extracts the CU block structure, while the IPM branch collects the IPM information. Then, CU block structure gradient and IPM mapping are performed to construct pixel-level CU block structure gradient map and IPM feature map, respectively. These two feature map are concatenated along the channel dimension and fed into GradIPMFormer network for joint structural modeling, which finally predicts whether the input frame is cover or stego.}
    \label{Framework}
\end{figure*}

Our proposed video steganalysis framework based on CU block structure gradients and IPM mapping is illustrated in Fig. \ref{Framework}. The overall pipeline consists of two core components: pixel-level feature map construction and joint structural modeling (GradIPMFormer network). First, the CU block structure and the IPM of each prediction unit are collected for the current frame. Then, pixel-level feature map construction is conducted in CU and IPM branches. Second, the CU block structure gradient map and IPM feature map are spatially aligned and concatenated to form a unified multi-channel representation. The fused feature map is then fed into the proposed GradIPMFormer network, which leverages Transformer-based global dependency modeling to capture subtle structural perturbations and discriminate between covers and stegos.

\subsection{Pixel-level Feature Map Construction} \label{III-B}

As discussed in Section \ref{intro}, CU block structure–based steganography inevitably alters the block structure, which in turn affects the IPM decision. Therefore, to more effectively capture steganographic traces, we propose a fused feature construction strategy. Specifically, the proposed representation consists of a CU block structure gradient map and an IPM feature map, enabling a more comprehensive characterization of embedding-induced structural and syntactic perturbations.

\subsubsection{CU Block Structure Gradient Map Construction} \label{III-B1}

\begin{figure}[htbp]
    \centering
    \includegraphics[width=1\linewidth]{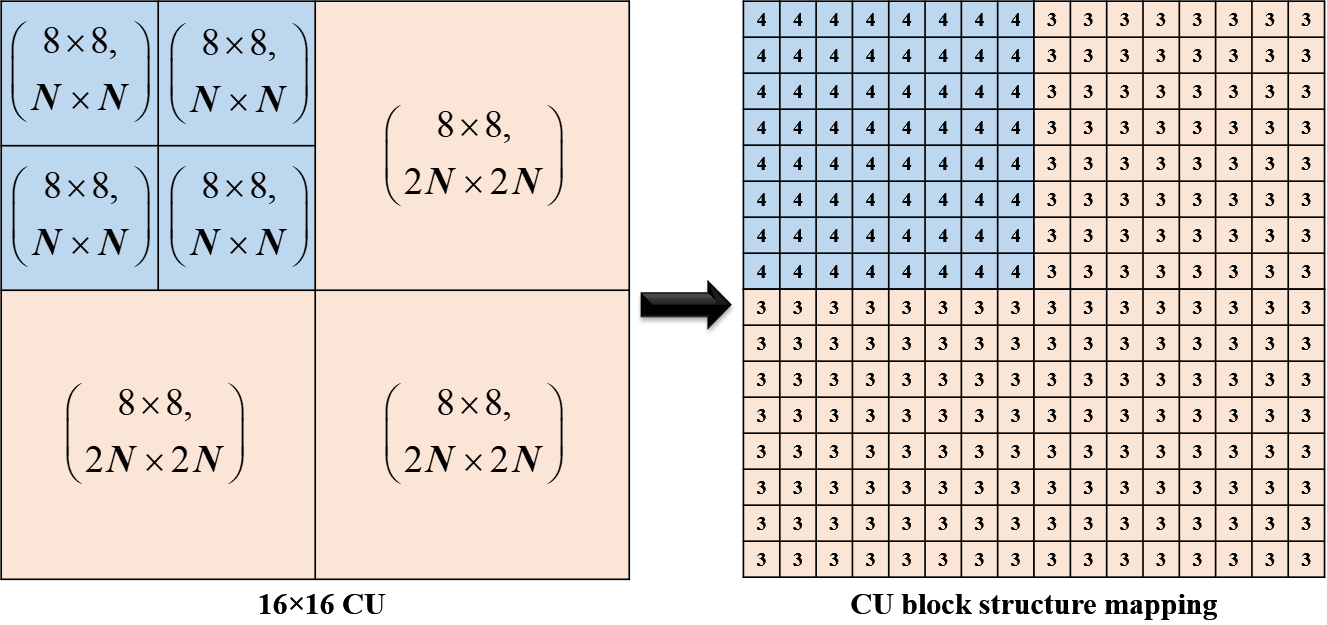}
    \caption{Example of CU block structure mapping construction for $16 \times 16$ CU}
    \label{CUmap}
\end{figure}

During H.265/HEVC encoding, each I frame is partitioned into multiple Coding Units (CUs) with different spatial sizes and hierarchical levels. Let the set of CUs in the current I frame $FR$ be denoted as:
\begin{equation}
CU = \{cu_1, cu_2, \ldots, cu_n\},
\end{equation}
where $n$ is the total number of CUs in frame $FR$. Suppose that the $k$-th CU of $FR$ is $cu_k$. To capture hierarchical partition information in a unified form, we first define the structural label of $cu_k$ as:
\begin{equation} 
BM_{cu_k} = 
\begin{cases} 
4, & size_{cu_k}=(8 \times 8, N \times N), \\ 
3, & size_{cu_k}=(8 \times 8, 2N\times2N), \\ 
2, & size_{cu_k}=(16 \times 16), \\ 
1, & size_{cu_k}=(32 \times 32), \\ 
0, & size_{cu_k}=(64 \times 64),
\end{cases} 
\end{equation}
where $size_{cu_k}$ denotes the block size and intra partition pattern of $cu_k$. Then, we obtain the width and height of $cu_k$ as $W$ and $H$, respectively, and fill $BM_{cu_k}$ into all pixel positions belonging to $cu_k$ to construct $cumap_{cu_k}$ as:
\begin{equation} 
 cumap_{cu_k}(i,j) = BM_{cu_k},
\end{equation}
where $0\le i<H$ and $0\le j <W$. After all CUs in the current frame are processed and placed according to their spatial positions, these local mappings jointly form the frame-level CU block structure mapping. Fig.~\ref{CUmap} shows an example of CU block structure mapping construction.

Then, to enhance sensitivity to structural discontinuities, first-order discrete differences are computed on the frame-level CU block structure mapping. The horizontal and vertical first-order discrete differences of $cu_k$ can be expressed as:
\begin{equation}
\Delta_x cumap_{cu_k}(i,j) =
cumap_{cu_k}(i,j) - cumap_{cu_k}(i,j+1), 
\end{equation}
\begin{equation}
\Delta_y cumap_{cu_k}(i,j) =
cumap_{cu_k}(i,j) - cumap_{cu_k}(i+1,j).
\end{equation}

If the neighboring position belongs to another CU, its value is taken from the corresponding mapped region of that adjacent CU. For the frame boundary, the corresponding difference is set to zero. Then, the magnitude of the CU block structure gradient is defined as:
\begin{equation}
g_{cu_k}(i,j)
=
\left| \Delta_x cumap_{cu_k}(i,j) \right|
+
\left| \Delta_y cumap_{cu_k}(i,j) \right|.
\end{equation}

Finally, all gradient responses are arranged according to their original spatial positions to form the complete frame-level CU block structure gradient map, denoted as $\mathbf{G}$.

\subsubsection{IPM Feature Map Construction} \label{III-B2}
In H.265/HEVC, IPM is determined at the Prediction Unit (PU) level, which means that each PU is associated with a unique prediction mode. 
Let the set of PUs in the current I frame $FR$ are:
\begin{equation}
    PU = \{pu_1, pu_2, \ldots, pu_m\},
\end{equation}
where $m$ represents the total number of PUs in $FR$. For the $k$-th prediction unit $pu_k$, with a size of $H \times W$ and an IPM denoted as $mode(pu_k)$, the pixel-level IPM mapping of $pu_k$ can be constructed as:
\begin{equation}
ipmmap_{pu_k}(i,j) = mode(pu_k),
\label{eq:ipm_map}
\end{equation}
where $ipmmap_{pu_k}(i,j)$ represents the IPM mapping value at position $(i,j)$ of $pu_k$, and $0\le i<H$ and $0\le j <W$.

\begin{figure}[htbp]
    \centering
    \includegraphics[width=0.92\linewidth]{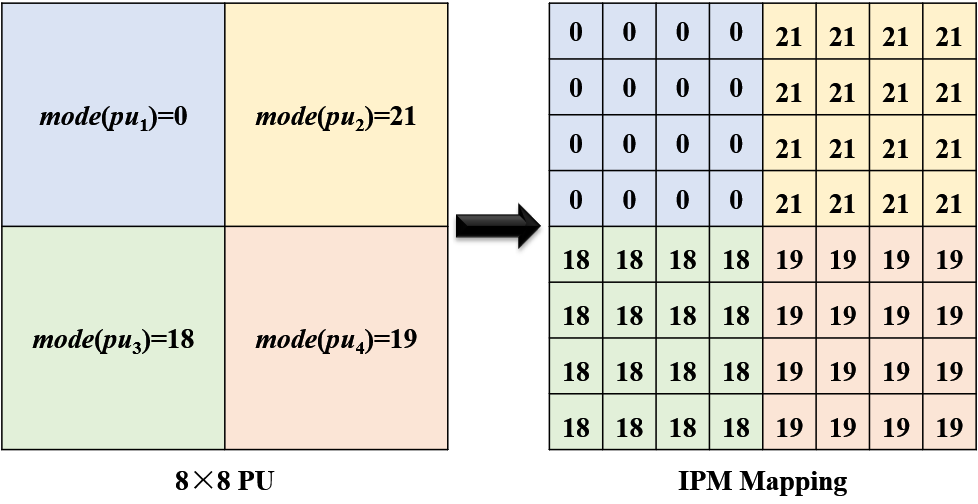}
    \caption{Example of IPM mapping construction for $8 \times 8$ PU}
    \label{IPMmap}
\end{figure}

All PUs in the current frame $FR$ are processed using (\ref{eq:ipm_map}), and their corresponding mappings are concatenated according to their spatial positions to form the complete pixel-level IPM mapping. Fig.~\ref{IPMmap} shows an example of IPM mapping construction for $8 \times 8$ PU.

Since IPM values are categorical without inherent order, we use one-hot encoding to represent each mode as an independent channel, avoiding incorrect numerical relationships and enabling better feature learning. For each prediction unit $pu_k$, the one-hot representation is defined as:
\begin{equation}
\mathbf{E}_{pu_k}(i,j) = [e_{pu_k}(0,i,j), \ldots, e_{pu_k}(34,i,j)],
\end{equation}
where
\begin{equation}
e_{pu_k}(t,i,j) =
\begin{cases}
1, & \text{if } mode(pu_k)=t, \\
0, & \text{otherwise}.
\end{cases}
\end{equation}

All one-hot IPM mappings are concatenated to form the 35-channel frame-level IPM (one-hot) feature map $\mathbf{E}$.


\subsubsection{Pixel-Level Alignment and Feature Fusion} \label{III-B3}
The CU block tructure gradient map and the IPM (one-hot) feature map are concatenated along the channel dimension to obtain:
\begin{equation}
\mathbf{F} = [\mathbf{G}, \mathbf{E}].
\end{equation}

The fused feature map $\mathbf{F}$ is then fed into the subsequent proposed GradIPMFormer network for joint modeling of structural partition patterns and directional prediction semantics.

\subsection{GradIPMFormer Network} \label{III-C}

\begin{figure*}
    \centering
    \includegraphics[width=\linewidth]{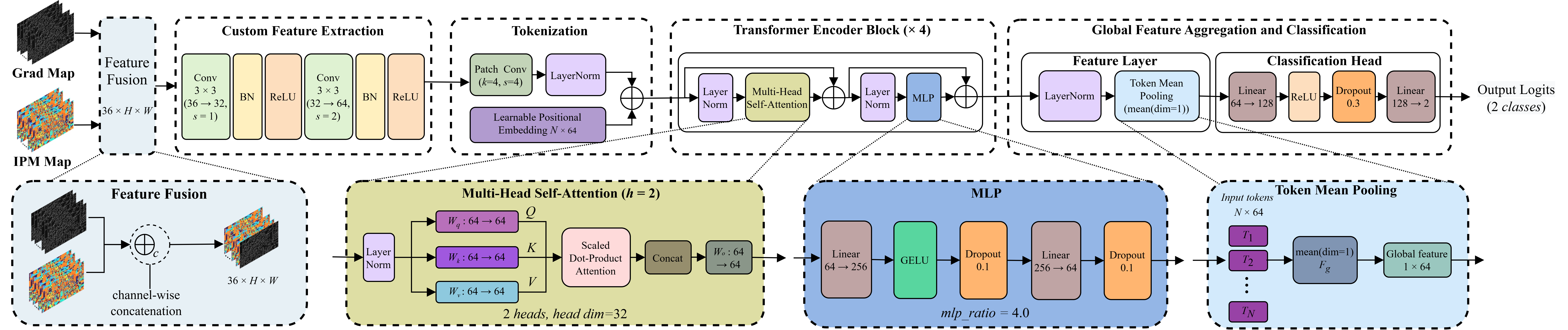}
    \caption{The overall framework of GradIPMFormer network}
    \label{fig:GradIPMFormer}
\end{figure*}

To effectively capture the complex perturbation patterns introduced by CU block structure-based steganography in the spatial domain, we design a Transformer-based steganalysis network with joint inputs of structural gradients and one-hot IPMs, termed \textit{GradIPMFormer}. The network consists of four key components: a custom feature extraction module, a tokenization module, a Transformer encoder block module, and a global feature aggregation and classification module. The overall architecture is illustrated in Fig.~\ref{fig:GradIPMFormer}.
The design of GradIPMFormer is motivated by the hierarchical and spatially structured nature of CU block structure-based steganography. Unlike natural image textures, the steganographic traces considered in this work are mainly reflected in local CU boundary variations, quadtree partition discontinuities, and their coupled changes with IPM distributions. Therefore, the network should not only capture fine-grained local boundary artifacts, but also model long-range dependencies among different CU regions. To this end, GradIPMFormer first uses a lightweight convolutional stem to preserve intra-block local structural details, and converts the extracted feature map into a sequence of structure-aware tokens for intra-block global modeling.

\subsubsection{Custom Feature Extraction Module} \label{III-C1}

Given that CU structural perturbations exhibit strong spatial locality, GradIPMFormer first employs a lightweight convolutional module to extract local features from the joint input tensor $\mathbf{F}$. This module consists of two $3 \times 3$ convolutional layers, where the second layer adopts a stride of 2 to downsample the feature map and enlarge the receptive field. The process is formulated as:
\begin{equation}
\mathbf{F}_{1}=\sigma\!\left(\mathrm{BN}\!\left(\mathrm{Conv}^{s=1}_{3\times 3}(\mathbf{F})\right)\right),
\label{eq:conv_embed_1}
\end{equation}
\begin{equation}
\mathbf{F}_{2}=\sigma\!\left(\mathrm{BN}\!\left(\mathrm{Conv}^{s=2}_{3\times 3}(\mathbf{F}_{1})\right)\right),
\label{eq:conv_embed_2}
\end{equation}
where $\mathrm{Conv}^{s=1}_{3\times 3}$ and $\mathrm{Conv}^{s=2}_{3\times 3}$ denote the 2D convolution with stride 1 and stride 2, respectively. $\mathrm{BN}(\cdot)$ denotes batch normalization, $\sigma(\cdot)$ is the ReLU activation function. In this module, the first convolution maps the 36-channel fused input to 32 channels, and the second convolution maps the feature dimension from 32 to 64 while reducing the spatial resolution by a factor of 2. This module enables the network to capture fine-grained local structural perturbations, providing representations for subsequent modeling.

\subsubsection{Tokenization Module} \label{III-C2}

A patch embedding strategy is adopted to convert the two-dimensional feature representation into the token sequence required by the Transformer. The input feature map is denoted as $\mathbf{F}_{2}\in\mathbb{R}^{C_s\times H_2\times W_2}$, where $C_s$ denotes the number of feature channels, and $H_2$ and $W_2$ denote the height and width of the feature map, respectively. A convolution operation with a kernel size of $P\times P$ and a stride of $P$ is applied to $\mathbf{F}_{2}$:

\begin{equation}
\mathbf{Z} = \mathrm{Conv}^{s=P}_{P\times P}(\mathbf{F}_{2}).
\label{eq}
\end{equation}

In our implementation, $P=4$. Since both the kernel size and stride are set to 4, each output position in $\mathbf{Z}$ is generated from a non-overlapping $4\times4$ local region of $\mathbf{F}_{2}$. Therefore, the spatial dimensions of $\mathbf{Z}$ are given by:

\begin{equation}
H_z=\frac{H_2}{P}=\frac{H_f}{2P}, \qquad
W_z=\frac{W_2}{P}=\frac{W_f}{2P}.
\label{eqz}
\end{equation}
where $H_f$ and $W_f$ denote the height and width of the original fused input feature map, when $P=4$, the spatial resolution of $\mathbf{Z}$ is reduced by a factor of 8 relative to the original input map. Each token in $\mathbf{Z}$ corresponds to an $8\times8$ spatial region in the original input coordinate system. Therefore, the constructed token grid has a spatial scale consistent with the minimum $8\times8$ CU granularity in H.265/HEVC. it can be seen that an individual token can represent a local CU-level structural pattern, while larger CUs and CTUs are jointly represented by multiple spatially adjacent tokens. In this way, the token sequence preserves the spatial organization of CU partitions while providing a compact representation for global dependency modeling. Then, a token sequence $\mathbf{T}$ with a length of $N$ are computed from the resulting feature tensor $\mathbf{Z}\in\mathbb{R}^{d\times H_z\times W_z}$ as:
\begin{equation}
\mathbf{T}
=
\mathrm{Transpose}\left(\mathrm{Flatten}(\mathbf{Z})\right)
=
[\mathbf{t}_{1}, \mathbf{t}_{2}, \ldots, \mathbf{t}_{N}],
\label{eq:token_seq}
\end{equation}
where $\mathrm{Transpose}(\cdot)$ and $\mathrm{Flatten}(\cdot)$ are transpose and flatten function, respectively. According to (\ref{eqz}), $N=H_zW_z=\frac{H_f}{2P}\times\frac{W_f}{2P}$ denotes the number of patch tokens, and each token $\mathbf{t}_{i}\in\mathbb{R}^{d}$, has an embedding dimension of $d=64$, and thus $\mathbf{T}\in\mathbb{R}^{N\times d}$. To mitigate channel-wise scale variations, Layer Normalization (LN) is applied to each token as $\hat{\mathbf{t}}_{i} = LN(\mathbf{t}_{i})$ to obtain the normalized token sequence:
\begin{equation}
\hat{\mathbf{T}} = [\hat{\mathbf{t}}_{1}, \hat{\mathbf{t}}_{2}, \ldots, \hat{\mathbf{t}}_{N}].
\label{eq:token_seq_norm}
\end{equation}

Since the Transformer encoder is permutation-invariant and lacks explicit spatial awareness, we introduce a learnable positional embedding sequence to encode the spatial layout of patches. Specifically, the positional embedding sequence is defined as:
\begin{equation}
\mathbf{E}^{pos}
=
[\mathbf{e}^{pos}_{1}, \mathbf{e}^{pos}_{2}, \ldots, \mathbf{e}^{pos}_{N}],
\label{eq:pos_embed}
\end{equation}
where $\mathbf{e}^{pos}_{i}$ denotes the learnable positional vector assigned to the $i$-th patch token. These positional vectors are defined as trainable parameters and optimized jointly with the entire network during training to encode the spatial order of different patch locations. The final input sequence to the Transformer encoder is obtained by adding the corresponding positional vector to each normalized token:
\begin{equation}
\mathbf{T}^{0}
=
\hat{\mathbf{T}} + \mathbf{E}^{pos}.
\label{eq:transformer_input}
\end{equation}

This design preserves local structural semantics while enabling the model to capture spatial relationships across different CU regions.

\subsubsection{Transformer Encoder Block Module} \label{III-C3}

The token sequence $\mathbf{T}^{0}$ is fed into $L$ stacked Transformer encoder blocks (described in Section~\ref{II-c}) to model long-range dependencies among CU-level regions. 
To describe this layer-wise encoding process, let $\mathrm{TFB}(\cdot)$ denote a Transformer encoder block consisting of multi-head self-attention, a feed-forward network, Layer Normalization, and residual connections. The token sequence is progressively updated as:
\begin{equation}
\mathbf{T}^{l}
=
\mathrm{TFB}\left(\mathbf{T}^{l-1}\right),
\quad l=1,2,\ldots,L,
\label{eq:transformer_update}
\end{equation}
where $\mathbf{T}^{l-1}$ and $\mathbf{T}^{l}$ denote the input and output token sequences of the $l$-th Transformer encoder block, respectively. Through this layer-wise transformation, each token progressively aggregates contextual information from other CU-level regions.
Through self-attention, GradIPMFormer models long-range dependencies among CU-level tokens, enabling it to capture non-local inconsistencies in quadtree partition patterns. Since each token integrates CU-gradient and IPM information, the Transformer encoder can further characterize the correlations between CU local boundary perturbations and prediction-mode drift that are difficult to capture using only local convolutional operations. After $L$ Transformer encoder blocks, the output token sequence is denoted as:
\begin{equation}
\mathbf{T}^{L}=[\mathbf{t}^{L}_{1},\mathbf{t}^{L}_{2},\ldots,\mathbf{t}^{L}_{N}].
\end{equation}

\subsubsection{Global Feature Aggregation and Classification Module} \label{III-C4}

To obtain a global representation, a final Layer Normalization is first applied to the output token sequence, followed by token mean pooling across the token dimension:
\begin{equation}
\tilde{\mathbf{T}}^{L}=\mathrm{LN}(\mathbf{T}^{L}),
\end{equation}
\begin{equation}
\mathbf{F}_{g} = \frac{1}{N} \sum_{i=1}^{N} \tilde{\mathbf{T}}^L(i),
\end{equation}
where $\mathbf{F}_{g}\in\mathbb{R}^{d}$ denotes the aggregated global feature representation. The aggregated feature $\mathbf{F}_{g}$ is then fed into a classification head $\mathrm{MLP}_{cls}(\cdot)$ consisting of a linear layer, a ReLU activation, a dropout layer, and a final linear layer. The classifier outputs two-dimensional logits for cover/stego prediction:
\begin{equation}
\mathbf{o} = \mathrm{MLP}_{cls}(\mathbf{F}_{g}),
\end{equation}
where $\mathbf{o}\in\mathbb{R}^{2}$ denotes the output logits. During inference, the class probability can be obtained by applying a Softmax operation to $\mathbf{o}$. Algorithm~\ref{alg:gradipmformer} summarizes the configuration and forward propagation process of GradIPMFormer.

\begin{algorithm}[t]
\caption{Configuration and Forward Propagation of GradIPMFormer}
\label{alg:gradipmformer}
\begin{algorithmic}[1]
\REQUIRE Fused feature map $\mathbf{F}\in\mathbb{R}^{C_{in}\times H_f\times W_f}$, where $C_{in}=36$; stem output channels $C_s=64$; patch size $P=4$; embedding dimension $D=64$; Transformer depth $L=4$; attention heads $h=2$; MLP expansion ratio $r=4$; classifier hidden dimension $D_c=128$.
\ENSURE Output logits $\mathbf{o}$.

\STATE $\mathbf{F}_1 \leftarrow \mathrm{ReLU}(\mathrm{BN}(\mathrm{Conv}_{3\times3}^{s=1}(\mathbf{F})))$
\STATE $\mathbf{F}_2 \leftarrow \mathrm{ReLU}(\mathrm{BN}(\mathrm{Conv}_{3\times3}^{s=2}(\mathbf{F}_1)))$

\STATE $\mathbf{Z} \leftarrow \mathrm{Conv}_{P\times P}^{s=P}(\mathbf{F}_2)$
\STATE $\mathbf{T} \leftarrow \mathrm{Transpose}(\mathrm{Flatten}(\mathbf{Z}))$
\STATE $\hat{\mathbf{T}} \leftarrow \mathrm{LN}(\mathbf{T})$

\STATE $\mathbf{T}^{0} \leftarrow \hat{\mathbf{T}}+\mathbf{E}_{pos}$

\FOR{$l=1$ to $L$}
    \STATE $\mathbf{T}^{l} \leftarrow \mathrm{TFB}(\mathbf{T}^{l-1})$
\ENDFOR

\STATE $\tilde{\mathbf{T}}^{L} \leftarrow \mathrm{LN}(\mathbf{T}^{L})$
\STATE $\mathbf{F}_g \leftarrow \frac{1}{N}\sum_{i=1}^{N}\tilde{\mathbf{T}}^{L}(i)$
\STATE $\mathbf{o} \leftarrow \mathrm{MLP}_{cls}(\mathbf{F}_g)$
\RETURN $\mathbf{o}$
\end{algorithmic}
\end{algorithm}

\section{Experimental Results and Analysis}\label{sec4}
\subsection{Experimental Setup} \label{IV-A}
All experiments in this paper were conducted on a machine equipped with an NVIDIA GeForce RTX 4090 GPU, running at 3.1 GHz with 24 GB of memory. The proposed steganalysis algorithm was implemented in Python 3.11.
\subsubsection{Video Dataset} \label{IV-A1}
The experiments use a video dataset constructed from 36 standard YUV video sequences, including 31 sequences with a resolution of $1920\times1080$ (1080P) and 5 sequences with a resolution of $832\times480$ (480P). The details of the dataset are shown in TABLE \ref{tab1}. All the YUV sequences are obtained from the publicly available Xiph.org video test media (\url{https://media.xiph.org/}).

\begin{table}[htbp]
\centering
\caption{YUV test sequences}
\label{tab1}
\renewcommand{\arraystretch}{1.1} 
\setlength{\tabcolsep}{4pt}       
\begin{tabular}{c c m{5.5cm}}  
\hline
\textbf{Index} & \textbf{Resolution} & \textbf{Sequence} \\
\hline
1 & 832 $\times$ 480 & BasketballDrill, BasketballDrillText, BQMall, PartyScene, RaceHorses \\[3pt]
2 & 1920 $\times$ 1080 & Aspen, BasketballDrive, BigBuckBunny, BlueSky, BQTerrace, Cactus, ControlledBurn, CrowdRun, Dinner, DucksTakeOff, ElephantsDream, Factory, InToTree, Kimono1, Life, OldTownCross, ParkJoy, ParkScene, PedestrianArea, RedKayak, Riverbed, RushFieldCuts, RushHour, SintelTrailer, SnowMnt, SpeedBag, Station2, Sunflower, TouchdownPass, Tractor, WestWindEasy \\
\hline
\end{tabular}
\end{table}

For the 1080P videos, a total of 760 subsequences are generated, with each subsequence containing 60 frames. For the 480P videos, 47 subsequences are produced, each also consisting of 60 frames. In total, the dataset consists of 48,420 video frames. All YUV sequences follow the 4:2:0 format. To improve efficiency, all steganography algorithms are implemented on the HM 16.15. The Group Of Pictures (GOP) structure is configured as “IPPPPPPPPPPP” for 1080P videos and “IPPP” for 480P videos. It is worth noting that the “IPPP” configuration is adopted for the 480P videos because the number of available 480P source videos is significantly smaller than that of 1080P. A shorter GOP structure is needed to generate a relatively adequate number of samples.

\subsubsection{Steganography Methods} \label{IV-A2}
To evaluate the detection performance of the proposed steganalysis algorithm, we employ four CU block structure-based steganography methods: Tew et al.~\cite{tew2014information} (denoted as Tar1), Dong et al.~\cite{dong2022adaptive} (denoted as Tar2), Yang et al.~\cite{yang2024quad} (denoted as Tar3), and Wang et al.~\cite{wang2024adaptive} (denoted as Tar4).

\subsubsection{Setups for Performance Evaluation} \label{IV-A3}
A total of five experimental setups are carefully designed to comprehensively evaluate the steganalysis performance, robustness, and cross-domain generalization of the proposed method.

\noindent\textbf{Setup 1: Steganalysis under different QP settings.} In this setting, we evaluate four representative H.265/HEVC steganography algorithms under the same quantization configuration. Specifically, for each algorithm, video samples are encoded with QP values of 26, 32, and 38, and the payload is set to 0.1 bpc, 0.3 bpc, and 0.5 bpc (bits per cover). For each steganography algorithm, a separate steganalysis detector is trained and tested on its corresponding dataset.

\noindent\textbf{Setup 2: Comparison with representative video steganalysis methods.} Since there are currently no publicly available and reproducible baselines specifically designed for CU block structure-based steganography, we select three representative H.265/HEVC video steganalysis methods as baselines, namely the methods of Cao et al.~\cite{cao2025steganalytic}, Sheng et al.~\cite{sheng2017prediction}, and Dai et al.~\cite{dai2023hevc}. These methods model abnormal variations of other H.265/HEVC syntax elements from different perspectives, providing references for evaluating the advantages of our proposed method.

\noindent\textbf{Setup 3: Comparison with different networks.} To further verify the effectiveness and transferability of the proposed feature extraction method and GradIPMFormer, we compare our approach with several representative networks, including NRNet~\cite{liu2020steganalysis}, PUNet~\cite{dai2023hevc} (the network used in Dai~\cite{dai2023hevc}), ZhangNet~\cite{zhang2021cnn}, and CENet~\cite{dai2025hevc}.  Specifically, the features extracted using the proposed feature extraction method were fed into each competing network for classification. All competing networks are evaluated under the same data splits and evaluation protocols to ensure fairness and reproducibility. 

\noindent\textbf{Setup 4: Robustness under mixed QP control and embedding conditions.} This setting aims to examine the robustness of the proposed steganalyzer under heterogeneous encoding and embedding conditions. In this Setup, the stego samples produced in Setup 1 under different QPs and embedding payloads are merged into a unified dataset, while keeping the 1080P and 480P samples in two separate groups. This setting evaluates whether the proposed method can maintain stable discriminative ability when the training data are no longer condition-uniform.

\noindent\textbf{Setup 5: Generalization under the Cover-Source Mismatch (CSM) setting.} CSM is one of the most critical factors hindering the practical deployment of steganalyzers. To simulate realistic detection scenarios, we train and test the steganalyzer using different video samples. Specifically, the training data consist of a mixture of 1080P video samples generated by Tar2 and Tar4 from Setting 1. The test data include 480P video samples generated by Tar1 and Tar3 from Setup 3. Therefore, there are significant differences between the training and test sets in terms of embedding methods and resolution.

\subsubsection{Training and Classification} \label{IV-A4}
For all setups (Setup 1–5), both cover and stego samples are randomly divided into training and testing sets at a ratio of 4:1. Meanwhile, a further 8:2 split is applied to the training set to form the training and validation subsets, which are used to monitor convergence and prevent overfitting. Training is conducted for 50 epochs using Adam optimizer (initial learning rate $1\times10^{-4}$, weight decay $1\times10^{-4}$). A ReduceLROnPlateau strategy is adopted to adjust the learning rate according to the validation loss. The loss function is the weighted cross-entropy, where class weights are dynamically computed based on the ratio of cover and stego samples to alleviate class imbalance. Automatic Mixed Precision (AMP) is employed to accelerate both forward and backward passes. The final performance is reported using the model checkpoint that achieves the best validation accuracy and is evaluated on the independent testing set.

\subsubsection{Performance Evaluation Index} \label{IV-A5}
The performance of the proposed steganalysis model is quantitatively assessed using standard evaluation metrics. In this study, classification accuracy is adopted as the primary metric to measure detection performance. The detection accuracy $P_{ACC}$ is defined as:
\begin{equation}
    P_{ACC} = \frac{T P+T N}{T P+T N+F P+F N},
\end{equation}
where $TP$, $TN$, $FP$, and $FN$ denote the numbers of true positives, true negatives, false positives, and false negatives. 

\subsection{Test Performances} \label{IV-B}
\subsubsection{Performance Evaluation of Steganalysis Under Different QP Settings} \label{IV-B-1}
TABLE~\ref{tab2} reports the detection accuracy of the proposed method on four representative H.265/HEVC steganographic algorithms (Tar1–Tar4) under different QP and payload settings. Overall, the detection accuracy shows a consistent upward trend as the payload increases from 0.1 to 0.5, reaching or even approaching 100\% in some configurations. This indicates that stronger embedding introduces more noticeable perturbations to the coding structure and intra prediction decisions, which can be effectively captured by the joint modeling of CU block structure gradients and IPM mapping. Across different QP settings, higher QP values generally yield better detection performance. It can be seen that the detection accuracy of Tar1–Tar4 improves to varying degrees when QP increases from 26 to 38 on both resolutions. This suggests that under strong compression, the encoder imposes stricter rate–distortion constraints, making steganographic embedding more likely to disrupt the original optimal CU structures and prediction modes, thereby producing more salient anomalies. In contrast, under low-QP conditions, higher coding redundancy makes structural perturbations relatively more concealed, increasing detection difficulty.

In terms of resolution, detection performance at 1080P is generally superior to that at 480P. High-resolution videos contain more CU blocks and richer hierarchical structural information, allowing embedding-induced perturbations to be more fully manifested in the spatial domain. Although the accuracy slightly decreases at 480P, it remains high in most configurations, demonstrating the robustness of the proposed method to resolution variations. Regarding different steganographic algorithms, Tar1 and Tar3 are more detectable in most settings, whereas Tar4 achieves relatively lower detection accuracy, particularly under low-QP and low-payload conditions. This suggests that Tar4 perturbs the coding structure more conservatively, resulting in less steganographic traces. Nevertheless, under medium-to-high payloads and high-QP settings, the proposed method still achieves good performance, further validating the effectiveness of our strategy in different steganographic conditions.

\begin{table}[htbp]
\caption{\small Detection performance ($P_{ACC}{\uparrow}$) under different QP and payload settings}
\label{tab2}
\centering
\small
\renewcommand{\arraystretch}{1.2}
\setlength{\tabcolsep}{5pt}
\begin{tabular}{@{}ccccccc@{}}
\toprule
\multirow{2}{*}{\textbf{Resolution}} & \multirow{2}{*}{\textbf{QP}} & \multirow{2}{*}{\textbf{Steganography}} & \multicolumn{3}{c}{\textbf{Payload (bpc)}} \\
\cmidrule(l){4-6}
 &  &  & \textbf{0.1} & \textbf{0.3} & \textbf{0.5} \\
\midrule

\multirow{12}{*}{1080P}
& \multirow{4}{*}{26}
& Tar1 & 89.67\% & 95.67\% & 100.00\% \\
&  & Tar2 & 85.72\% & 90.66\% & 96.58\% \\
&  & Tar3 & 88.82\% & 92.28\% & 99.80\% \\
&  & Tar4 & 80.70\% & 86.32\% & 90.39\% \\
\cmidrule(l){2-6}

& \multirow{4}{*}{32}
& Tar1 & 95.04\% & 99.61\% & 100.00\% \\
&  & Tar2 & 92.22\% & 95.24\% & 97.11\% \\
&  & Tar3 & 94.47\% & 96.05\% & 99.87\% \\
&  & Tar4 & 86.46\% & 90.38\% & 95.11\% \\
\cmidrule(l){2-6}

& \multirow{4}{*}{38}
& Tar1 & 97.37\% & 100.00\% & 100.00\% \\
&  & Tar2 & 95.59\% & 97.63\% & 97.76\% \\
&  & Tar3 & 96.34\% & 99.87\% & 99.65\% \\
&  & Tar4 & 92.78\% & 95.30\% & 97.37\% \\

\midrule
\multirow{12}{*}{480P}
& \multirow{4}{*}{26}
& Tar1 & 88.48\% & 93.43\% & 95.24\% \\
&  & Tar2 & 82.62\% & 86.43\% & 89.71\% \\
&  & Tar3 & 84.29\% & 86.19\% & 89.10\% \\
&  & Tar4 & 76.67\% & 81.90\% & 85.76\% \\
\cmidrule(l){2-6}

& \multirow{4}{*}{32}
& Tar1 & 92.52\% & 95.38\% & 96.19\% \\
&  & Tar2 & 85.38\% & 90.95\% & 93.05\% \\
&  & Tar3 & 88.29\% & 90.48\% & 93.67\% \\
&  & Tar4 & 83.81\% & 86.19\% & 90.71\% \\
\cmidrule(l){2-6}

& \multirow{4}{*}{38}
& Tar1 & 95.92\% & 97.92\% & 100.00\% \\
&  & Tar2 & 89.05\% & 93.67\% & 96.67\% \\
&  & Tar3 & 90.67\% & 93.48\% & 96.48\% \\
&  & Tar4 & 87.62\% & 90.71\% & 95.10\% \\

\bottomrule
\end{tabular}
\end{table}

\subsubsection{Comparison with Representative Video Steganalysis Methods} \label{IV-B-2}

\begin{table}[t]
\centering
\caption{Detection Performance ($P_{ACC}{\uparrow}$) comparison of different steganalysis methods at QP=32.}
\label{tab:setting2_qp32}
\setlength{\tabcolsep}{6pt}
\renewcommand{\arraystretch}{1.18}
\begin{tabular}{c c cccc}
\toprule
\multirow{2}{*}{\textbf{Resolution}} & \multirow{2}{*}{\textbf{Steganography}} & \multirow{2}{*}{\textbf{Method}} & \multicolumn{3}{c}{\textbf{Payload (bpc)}} \\
\cmidrule(lr){4-6}
& & & \textbf{0.1} & \textbf{0.3} & \textbf{0.5} \\
\midrule

\multirow{16}{*}{1080P}
& \multirow{4}{*}{Tar1}
& Cao~\cite{cao2025steganalytic}   & 51.08 & 52.26 & 53.11 \\
& & Sheng~\cite{sheng2017prediction} & 49.94 & 50.73 & 51.22 \\
& & Dai~\cite{dai2023hevc}   & 51.46 & 52.94 & 53.48 \\
& & Proposed   & \textbf{95.04} & \textbf{99.61} & \textbf{100.00} \\
\cmidrule(lr){2-6}

& \multirow{4}{*}{Tar2}
& Cao~\cite{cao2025steganalytic}   & 50.84 & 51.37 & 52.11 \\
& & Sheng~\cite{sheng2017prediction} & 49.76 & 50.18 & 50.93 \\
& & Dai~\cite{dai2023hevc}   & 51.12 & 52.04 & 52.68 \\
& & Proposed   & \textbf{92.22} & \textbf{95.24} & \textbf{97.11} \\
\cmidrule(lr){2-6}

& \multirow{4}{*}{Tar3}
& Cao~\cite{cao2025steganalytic}   & 50.96 & 51.58 & 52.74 \\
& & Sheng~\cite{sheng2017prediction} & 49.81 & 50.24 & 50.88 \\
& & Dai~\cite{dai2023hevc}   & 51.28 & 52.01 & 53.16 \\
& & Proposed   & \textbf{94.47} & \textbf{96.05} & \textbf{99.87} \\
\cmidrule(lr){2-6}

& \multirow{4}{*}{Tar4}
& Cao~\cite{cao2025steganalytic}   & 50.27 & 50.91 & 51.76 \\
& & Sheng~\cite{sheng2017prediction} & 49.15 & 49.86 & 50.63 \\
& & Dai~\cite{dai2023hevc}   & 50.84 & 51.25 & 52.12 \\
& & Proposed   & \textbf{86.46} & \textbf{90.38} & \textbf{95.11} \\
\midrule

\multirow{16}{*}{480P}
& \multirow{4}{*}{Tar1}
& Cao~\cite{cao2025steganalytic}   & 50.33 & 51.12 & 51.84 \\
& & Sheng~\cite{sheng2017prediction} & 49.21 & 49.96 & 50.43 \\
& & Dai~\cite{dai2023hevc}   & 50.87 & 51.55 & 52.07 \\
& & Proposed   & \textbf{92.52} & \textbf{95.38} & \textbf{96.19} \\
\cmidrule(lr){2-6}

& \multirow{4}{*}{Tar2}
& Cao~\cite{cao2025steganalytic}   & 49.95 & 50.62 & 51.28 \\
& & Sheng~\cite{sheng2017prediction} & 48.87 & 49.58 & 50.11 \\
& & Dai~\cite{dai2023hevc}   & 50.41 & 51.06 & 51.73 \\
& & Proposed   & \textbf{85.38} & \textbf{90.95} & \textbf{93.05} \\
\cmidrule(lr){2-6}

& \multirow{4}{*}{Tar3}
& Cao~\cite{cao2025steganalytic}   & 49.87 & 50.42 & 51.19 \\
& & Sheng~\cite{sheng2017prediction} & 48.93 & 49.64 & 50.07 \\
& & Dai~\cite{dai2023hevc}   & 50.33 & 50.91 & 51.62 \\
& & Proposed   & \textbf{88.29} & \textbf{90.48} & \textbf{93.67} \\
\cmidrule(lr){2-6}

& \multirow{4}{*}{Tar4}
& Cao~\cite{cao2025steganalytic}   & 49.68 & 50.14 & 50.96 \\
& & Sheng~\cite{sheng2017prediction} & 48.54 & 49.12 & 49.88 \\
& & Dai~\cite{dai2023hevc}   & 50.02 & 50.77 & 51.34 \\
& & Proposed   & \textbf{83.81} & \textbf{86.19} & \textbf{90.71} \\
\bottomrule
\end{tabular}
\end{table}

Since there is currently no steganalysis specifically for CU block structures, we compare our method with three representative H.265/HEVC video steganalysis methods, namely, Cao~\cite{cao2025steganalytic}, Sheng~\cite{sheng2017prediction}, and Dai~\cite{dai2023hevc}. They primarily detect variations in other syntax elements. This setting aims to evaluate the applicability of different methods under structured embedding perturbations from a more general perspective. As shown in TABLE~\ref{tab:setting2_qp32}, we conduct a comparison on four CU block structure-based steganography algorithms (Tar1–Tar4) under QP=32, across different payloads and two resolutions. Overall, the three comparative steganalysis methods have difficulty consistently capturing structural anomalies caused by changes in CU block structure and their coupling with the prediction process. In contrast, our method, by jointly modeling CU structural gradients and IPM mapping, can more comprehensively characterize the local structural perturbations and cross-block correlations introduced by CU block structure-based steganography, thereby exhibiting more stable and stronger detection capability under different resolutions and payload settings.

\subsubsection{Comparison with Different Networks} \label{IV-B-3}

\begin{figure*}
    \centering
    \includegraphics[width=1\linewidth]{final_network_comparison_multiQP.png}
    \caption{Detection accuracies ($P_{ACC}\uparrow$) of different networks against Tar1–Tar4 under different resolution and QP settings. A, B, C, D, E, and F indicate (480P, 26), (480P, 32), (480P, 38), (1080P, 26), (1080P, 32), and (1080P, 38), respectively.}
    \label{fig:setting3}
\end{figure*}

In this section, the effectiveness of the proposed method from the perspective of network architectures are evaluated. Specifically, We compare our networks with four representative deep steganalysis networks, including NRNet~\cite{liu2020steganalysis}, PUNet~\cite{dai2023hevc}, ZhangNet~\cite{zhang2021cnn}, and CENet~\cite{dai2025hevc}. It should be noted that Dai~\cite{dai2023hevc} in Section~\ref{IV-B-2} denotes the complete steganalysis method proposed in that work, where the original input features and detection pipeline are adopted. In contrast, PUNet~\cite{dai2023hevc} in this subsection refers only to the network architecture used in Dai~\cite{dai2023hevc}. For a fair comparison, all the compared networks are trained using the same feature representation as our method, rather than their original hand-crafted or syntax-specific inputs, and they are trained and tested under the same data splits and evaluation protocols. The detection accuracies are reported under different combinations of resolution and quantization parameters, which are shown in Fig.~\ref{fig:setting3}. In the figure, A–F correspond to (480P, 26), (480P, 32), (480P, 38), (1080P, 26), (1080P, 32), and (1080P, 38).

From the overall trend, all networks achieve detection accuracies above 60\% (and above 70\% in most cases) under different QP settings and resolutions, demonstrating that the proposed joint features provide effective structural cues for steganalysis. As the payload increases from 0.1 to 0.5, the detection performance of all networks generally improves across the four steganography algorithms. However, the performance varies among different networks under the same configuration. In particular, under low-payload, some comparative networks exhibit more pronounced performance degradation, suggesting that those simple models relying mainly on a single representation or local features may have difficulty consistently characterizing structural perturbations in CU block structure-based steganography. In contrast, GradIPMFormer maintains more stable advantages across most configurations on Tar1–Tar4, especially under low payload. This can be attributed to its enhanced ability to model long-range structural dependencies across CUs, which improves its robustness performance.

\subsubsection{Performance Evaluation of Mixed QP control and embedding conditions}\label{IV-B-4}


\begin{table*}[t]
\centering
\caption{Detection accuracies ($P_{ACC}{\uparrow}$) comparison of different steganalysis networks against Tar1--Tar4 with different QP and embedding strengths under mixed QP control and embedding conditions.}
\label{tab:setting4_comparison}
\setlength{\tabcolsep}{5pt}
\renewcommand{\arraystretch}{1.18}
\begin{tabular}{c ccccc !{\vrule width 0.8pt} ccccc}
\toprule
\multirow{2}{*}{\shortstack{\textbf{Target}\\\textbf{Methods}}}
& \multicolumn{5}{c}{\textbf{480P}}
& \multicolumn{5}{c}{\textbf{1080P}} \\
\cdashline{2-6}[1.5pt/1pt]\cdashline{7-11}[1.5pt/1pt]
& \textbf{ZhangNet~\cite{zhang2021cnn}} & \textbf{NRNet~\cite{liu2020steganalysis}} & \textbf{PUNet~\cite{dai2023hevc}} & \textbf{CENet~\cite{dai2025hevc}} & \textbf{Proposed}
& \textbf{ZhangNet~\cite{zhang2021cnn}} & \textbf{NRNet~\cite{liu2020steganalysis}} & \textbf{PUNet~\cite{dai2023hevc}} & \textbf{CENet~\cite{dai2025hevc}} & \textbf{Proposed} \\
\midrule
Tar1 & 85.24 & 92.10 & 94.79 & 95.10 & \textbf{100.00}
     & 91.74 & 95.58 & 96.45 & 99.84 & \textbf{100.00} \\
\cdashline{1-11}[1.5pt/1pt]
Tar2 & 79.49 & 82.20 & 84.06 & 85.20 & \textbf{95.66}
     & 84.55 & 85.77 & 87.22 & 90.03 & \textbf{98.15} \\
\cdashline{1-11}[1.5pt/1pt]
Tar3 & 78.30 & 80.46 & 90.77 & 87.08 & \textbf{96.11}
     & 85.16 & 88.58 & 91.11 & 93.07 & \textbf{99.63} \\
\cdashline{1-11}[1.5pt/1pt]
Tar4 & 65.65 & 77.08 & 79.37 & 80.46 & \textbf{94.35}
     & 76.77 & 79.30 & 81.41 & 83.73 & \textbf{96.25} \\
\bottomrule
\end{tabular}
\end{table*}

As shown in TABLE~\ref{tab:setting4_comparison}, we compare the proposed method with the four representative deep steganalysis networks under mixed QP control and embedding conditions, and report the detection accuracies on four CU block structure-based steganography algorithms (Tar1–Tar4) at both 480P and 1080P resolutions. Overall, the detection performance of the comparative networks is affected to varying degrees, and the impact is more evident in the more challenging Tar2–Tar4 scenarios. In contrast, the proposed method maintains more pronounced advantages across both resolutions and all four steganography algorithms, indicating stronger adaptability and stability under mixed coding configurations and mixed embedding strengths. Since the compared networks are fed with the same input representations, the superior performance should be attributed to the architectural design of GradIPMFormer. Compared with conventional CNN-based steganalysis networks, GradIPMFormer combines local feature extraction with Transformer-based global dependency modeling, enabling it to better characterize both fine-grained structural perturbations and their long-range contextual correlations. As a result, it can more effectively exploit the shared CU- and IPM-related evidence under mixed QP settings and embedding strengths, thereby achieving more stable and robust detection performance across different target steganography.

\subsubsection{CSM Steganalysis Evaluation Performance}\label{IV-B-5}
\begin{figure}
    \centering
    \includegraphics[width=\linewidth]{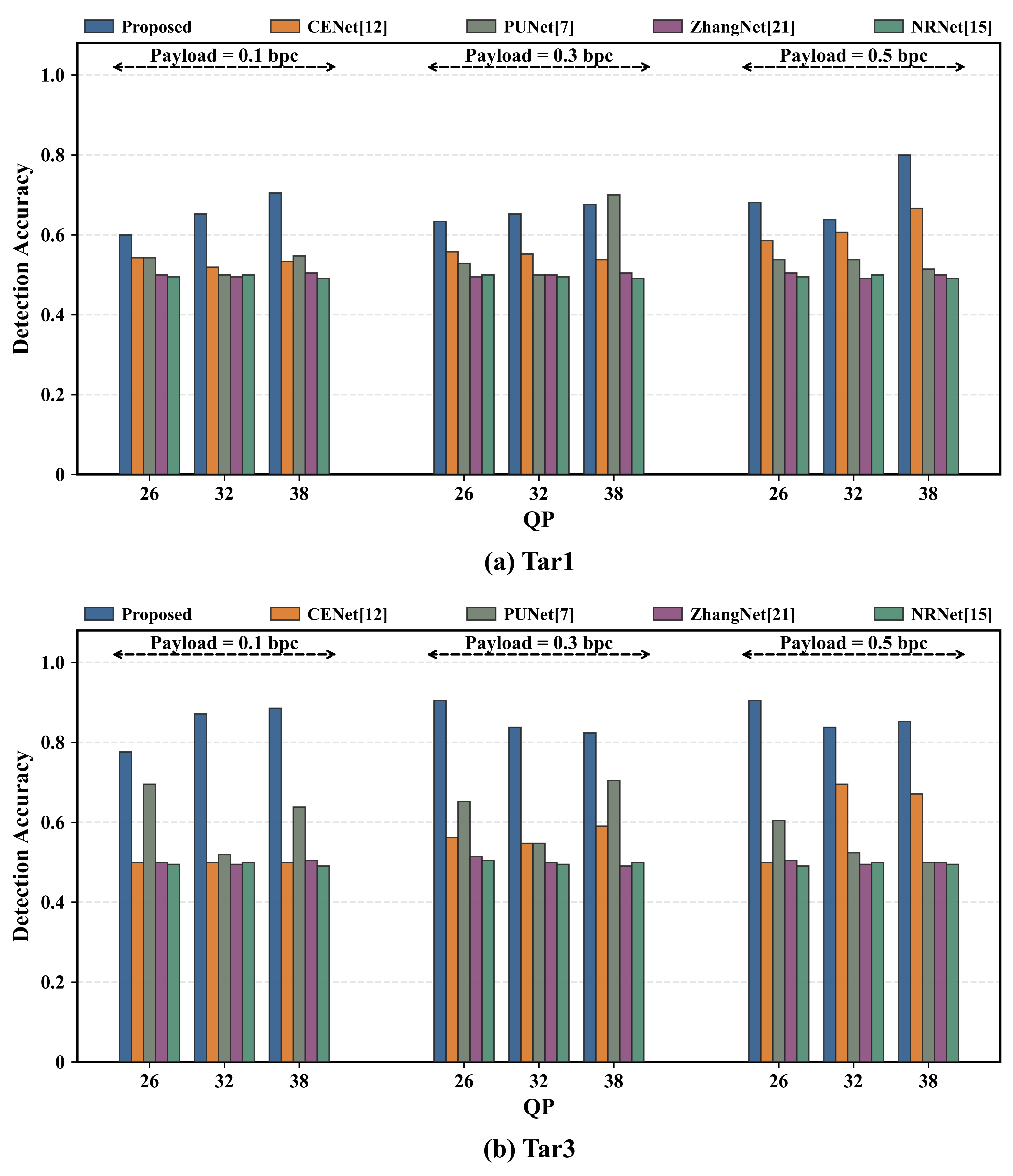}
    \caption{Detection accuracies($P_{ACC}{\uparrow}$) under the Cover-Source Mismatch (CSM) setting. }
    \label{fig:setting5}
\end{figure}

In this section, we evaluate the generalization capability of the proposed method under the cover-source mismatch (CSM) setting. This setting is designed to simulate a more realistic deployment scenario, where the training and test data differ simultaneously in both video resolution and embedding algorithms, thereby significantly increasing the detection difficulty. Specifically, the training set consists of 1080P stego videos generated by Tar2 and Tar4 from Setup 1, while the test set contains 480P stego videos generated by Tar1 and Tar3 from Setting 3. Consequently, the detector must handle a severe dual-domain shift caused by differences in both coding characteristics and steganographic distributions.

As shown in Fig.~\ref{fig:setting5}, we compare the detection accuracies of CENet~\cite{dai2025hevc}, PUNet~\cite{dai2023hevc}, ZhangNet~\cite{zhang2021cnn}, NRNet~\cite{liu2020steganalysis}, and the proposed network under different QPs and payloads. Overall, the CSM setting leads to noticeable performance degradation and instability for the competing methods. In particular, ZhangNet~\cite{zhang2021cnn} and NRNet~\cite{liu2020steganalysis} remain close to random guessing in most cases, indicating limited robustness to cross-domain distribution shifts. Although CENet~\cite{dai2025hevc} and PUNet~\cite{dai2023hevc} achieve relatively better results under several settings, their performances fluctuate significantly across different payloads and QPs, especially under low-payload conditions where the mismatch between training and testing distributions becomes more pronounced. In contrast, the proposed method maintains consistently superior performance, and the detection accuracy generally improves as the payload increases, which exhibits stronger robustness against distribution mismatch.

\subsection{Ablation Study} \label{IV-C}

\subsubsection{Effectiveness of CU Block Structure Gradient and IPM Mapping Representation}

To evaluate the effectiveness of the proposed CU block structure gradient representation and IPM mapping representation, we first conduct an ablation study by comparing three different input settings: Raw CU block structure features, CU block structure gradient features, and CU block structure gradient with IPM mapping features. The results are reported in TABLE~\ref{tab:cu_gradient_ablation}. In the table, \emph{Block} denotes the raw CU block structure feature without gradient transformation, while \emph{Gradients} denotes the CU block structure gradient feature. In contrast, \emph{Fused} represents our full input feature scheme, which combines CU block structure gradients with IPM mapping features. All experiments are conducted under the fixed setting of 1080P resolution, QP=32, and payload=0.3. 

\begin{table}[htbp]
\centering
\caption{Detection accuracies ($P_{ACC}{\uparrow}$, \%) of different representations at 1080P resolution (QP=32, payload=0.3).}
\label{tab:cu_gradient_ablation}
\setlength{\tabcolsep}{6pt}
\renewcommand{\arraystretch}{1.1}
\begin{tabular}{cccc}
\toprule
\textbf{Steganography} 
& \textbf{\emph{Block}} 
& \textbf{\emph{Gradients}} 
& \textbf{\emph{Fused}} \\
\midrule
Tar1 & 93.84 & 95.79 & \textbf{99.61} \\
Tar2 & 87.89 & 90.63 & \textbf{95.24} \\
Tar3 & 89.87 & 93.93 & \textbf{96.05} \\
Tar4 & 85.37 & 88.93 & \textbf{90.38} \\
\bottomrule
\end{tabular}
\end{table}

As shown in TABLE~\ref{tab:cu_gradient_ablation}, replacing the raw CU block feature with the CU block structure gradient representation consistently improves the detection performance across all four steganography methods. This demonstrates that gradient transformation can better emphasize local structural variations, boundary transitions, and block-partition perturbations caused by steganographic modification, whereas the raw CU block representation mainly describes the spatial distribution of CU sizes and may contain redundant smooth structural regions.

Furthermore, the proposed fused representation achieves the best performance on all target methods. Compared with CU Structure Gradients alone, the proposed method further improves the detection accuracy, indicating that IPM information provides complementary prediction-direction cues to the CU structural gradient features. Therefore, the combination of CU structure gradients and IPM mapping features can jointly capture both block-partition perturbations and prediction-mode changes, leading to more discriminative steganalysis features.
\subsubsection{Effectiveness of One-Hot Encoding}

After validating the effectiveness of CU structure gradient representation, we further investigate the impact of one-hot encoding of IPM mapping on detection performance by comparing two schemes: IPM without one-hot encoding (No one-hot) and IPM with one-hot encoding (one-hot). To control variables, we conduct evaluations under the same fixed setting of 1080P resolution, QP=32, and payload=0.3, and report detection accuracies for four CU block structure-based steganography algorithms (Tar1--Tar4), as shown in TABLE~\ref{tab:onehot_ablation}.

\begin{table}[htbp]
\centering
\caption{Detection accuracies ($P_{ACC}{\uparrow}$, \%) of the ablation study on one-hot encoding of IPM mapping at 1080P resolution (QP=32, payload=0.3).}
\label{tab:onehot_ablation}
\setlength{\tabcolsep}{10pt}
\renewcommand{\arraystretch}{1.1}
\begin{tabular}{ccc}
\toprule
\textbf{Steganography} & \textbf{No one-hot} & \textbf{one-hot} \\
\midrule
Tar1 & 97.43 & \textbf{99.61} \\
Tar2 & 92.06 & \textbf{95.24} \\
Tar3 & 95.38 & \textbf{96.05} \\
Tar4 & 87.11 & \textbf{90.38} \\
\bottomrule
\end{tabular}
\end{table}

Overall, one-hot encoding yields consistent performance improvements across all four steganography algorithms. This is mainly because the IPM index is inherently a discrete categorical variable. Directly feeding it as a numerical value may introduce an unnecessary implicit ordinal relationship, which can interfere with the modeling of distributional differences among prediction modes. In contrast, the one-hot representation characterizes mode information in a categorical form, making it easier for the network to learn discriminative patterns related to structural and directional perturbations.

\section{Conclusion}\label{sec5}
This paper addresses the limitations of existing video steganalysis methods that are insufficient for capturing CU-level structural perturbations. We propose a novel framework that jointly models CU block structure gradients and IPM mapping, where a structural-gradient representation is constructed to encode partition boundaries and fused with CU-aligned IPM features to characterize structure-coupled embedding artifacts. Based on this representation, we develop a Transformer-based model, GradIPMFormer, which effectively integrates local convolutional features with global self-attention to capture long-range structural dependencies. Experimental results demonstrate that the proposed method achieves superior and stable performance across diverse quantization parameters, resolutions, and steganographic scenarios, while maintaining strong robustness under mixed conditions and cover-source mismatch. Future work will explore finer structure-aware representations and improved domain generalization to enhance applicability in real-world settings.

\bibliographystyle{ieeetr}
\bibliography{BibTex}

\end{document}